\documentclass[11pt]{article}
\usepackage{amsmath}
\usepackage{epsfig}
\usepackage{amsfonts}

\def\d{\partial}

\def\cD{{\cal D}}

\def\m{\mu}
\def\n{\nu}

\def\t{\tau}

\def\~{\widetilde}

\def\bY3{\bar Y_{,3}}
\def\Y3{Y_{,3}}

\def\Y{{\bar Y}}

\def\`{\dot}
\def\be{\begin{equation}}
\def\ee{\end{equation}}
\def\bea{\begin{eqnarray}}
\def\eea{\end{eqnarray}}

\def\fn{\footnote}

\def\mn{{\mu\nu}}

\begin{document}

\title{\textbf{Stability of the lepton bag model based on the Kerr-Newman solution}}

\author{A. Burinskii \\
    Lab. of Theor. Phys., NSI Russian Academy of
Sciences,\\ B. Tulskaya 52  Moscow, 115191 Russia, bur@ibrae.ac.ru}

\maketitle

\begin{abstract}We show that the considered in  previous paper \cite{BurGrBag} lepton bag model, generating the
external gravitational and electromagnetic fields of the Kerr-Newman (KN) solution is
supersymmetric and represents a BPS-saturated soliton, interpolating between  internal vacuum state
and external  KN solution. We obtain Bogomolnyi equations for this phase transition, and show that
Bogomolnyi bound determines all important features of this bag model, including its stable shape.
 In particular, for stationary KN solution the BPS-bound provides stability of the ellipsoidal form
 of the bag and formation of the ring-string structure at its border, while for the periodic electromagnetic
 excitations of the KN solution, the BPS-bound controls deformation of the surface of the bag,
 reproducing the known flexibility of bag models.
  \end{abstract}

\newpage
\section{Introduction and overview} This paper is extension of the work \cite{BurGrBag}, in which we considered  the gravitating lepton
bag model based on the Kerr-Newman (KN) black hole solution.

 It has been discussed long ago that black holes  may be connected with elementary particles.
 However, the spin/mass ratio of elementary particles is extreme large, and the corresponding black hole loses the
 horizons, turning into an ultra-extreme (over-charged and over-rotating) KN solution with a naked singular
 ring, which forms a topological defect of space-time. As usual, emergence of  singularity is a hint for generalization of the theory,
and the Kerr singular ring created the problem of source of the KN solution. This problem proved to be very complicated,  and this year we can mark  the 50th anniversary of its discussions. Earlier attempts to build a source of the KN solution where discussed
 by Israel in \cite{Isr}, and Israel referred to the paper by Newman and Janis 1965 \cite{NewJan}, wherein nontriviality of this problem was at first indicated. Carter  obtained in \cite{Car} that the KN solution has gyromagnetic ratio $g=2 ,$ corresponding to that of the Dirac electron, and starting from this fact, Israel suggested in \cite{Isr} a classical model of the electron based on a rotating disklike source of the KN solution, enclosed by the Kerr singular ring.

 The consistent \emph{regular} model of the KN source was suggested by L\'opez, who built the KN source as a rotating vacuum bubble, covering the Kerr singular ring. In the same time, many properties of the KN source indicated its close relationships
with string models \cite{Bur0,IvBur,BurSen,Bur3Q},  and resolution of this duality was coming from the disklike soliton model \cite{BurSol}, in which the vacuum internal state of the L\'opez bubble-source was replaced by a superconducting false-vacuum formed by the Higgs mechanism of symmetry breaking. The ring-string  emerged in this model as a narrow tube of the electromagnetic (EM) potential concentrated at the sharp border of the disklike source, similar to the well-known Nielsen-Olesen vortex string model in the Landau-Ginzburg theory \cite{NO}.

Recently, this model was generalized to a gravitating bag model \cite{BurGrBag},  for which one of the known features is a flexibility and ability to create string-like structures \fn{Extended  particle-like soliton models based on the Higgs mechanism of symmetry breaking, such
as Q-balls, skirmions, bags and vortex strings, are widely discussed now. Flexibility of the bag models is used, in particular, for the flux-tube string models \cite{MIT,SLAC}.}

Principal peculiarity of the model considered in \cite{BurGrBag} was the requirement to
retain the external gravitational electromagnetic field of the Kerr-Newman (KN) solution, which as is
known \cite{Car,DKS} has gyromagnetic ratio $g=2 ,$ corresponding to that of the Dirac electron.
Such a bag may be considered as a semiclassical model for some particles of the electroweak sector of Standard Model,
 such as electron or muon, since the external gravitational and electromagnetic field of
these particles corresponds to KN solution with very good precision.

In this paper we show that this bag model is
supersymmetric and represents a BPS-saturated soliton, interpolating between a supersymmetric pseudo-vacuum state inside the bag and the external field of the exact KN solution. We obtain that all the considered in \cite{BurGrBag} important features
of this soliton follow unambiguously from the Bogomolnyi equations corresponding to
the BPS-saturated solution.

\subsection{Source of the KN solution as a spinning soliton}.
\noindent The Kerr-Schild form of KN metric is \cite{DKS} \be g_\mn =\eta_\mn + 2H k_\m k_\n , \label{gKS}\ee where $ \eta_\mn $ is metric of auxiliary Minkowski space\fn{We use signature $(- + + +)$.}  $M^4 ,$ and
\be H=\frac {mr -e^2/2}{r^2+a^2 \cos ^2 \theta}\label{HKN} \ee
 is a scalar function,  $r$ and
$\theta$ are ellipsoidal coordinates, and $ k_\m $ is the null vector field, $ k_\m k^\m =0 ,$ forming
 the Principal Null Congruence (PNC) $\cal K ,$ a vortex polarization of Kerr space-time.  The
surface $r=0$ represents a disk-  like "door" from negative sheet $r<0$ to positive one $r>0$. The
smooth extension of the solution from retarded to advanced sheet (together with smooth extension of
the Kerr PNC) occurs   via disk $ r=0 $ spanned by the Kerr singular ring $ r=0, \ \cos\theta=0 $
(see fig.1), and creates another PNC on the negative sheet.
The null vector fields $k^{\m\pm}(x)$ turns out to be different on these sheets, and
two different null congruences ${\cal K}^\pm ,$ create two different metrics $ g_\mn^\pm =\eta_\mn
+ 2H k_\m^\pm k_\n^\pm $ on the same Minkowski background.

  The mysterious two-sheeted structure of the Kerr geometry caused
  searching diverse models for source of the KN solution avoiding negative sheet.
  Relevant "regularization" of this space was suggested by L\'opez \cite{Lop}, who excised singular
  region together with negative sheet and replace it by a regular core with a flat internal metric $\eta_\mn .$
  The resulting \emph{vacuum bubble} should be matched with external KN solution along the boundary $r=R,$
  determined by the condition
  \be
H|_{r=R}(r)=0 \label{HR0} ,\ee which in accordance with (\ref{gKS}) and (\ref{HKN}) leads to
\be R=r_e = \frac {e^2}{2m} . \label{Hre}
\ee
 Since $r$ is Kerr's oblate radial coordinate, see Fig.2, the bubble-source takes an ellipsoidal form and covers the Kerr singular region, forming a flat space inside the disk of radius $r_c \sim a = \hbar/mc $  and thickness $r_e ,$ so that the degree of
 flatness $r_e/r_c \sim e^2 = \alpha \sim 137^{-1}$ corresponds to the fine structure constant.
\begin{figure}[ht]
\centerline{\epsfig{figure=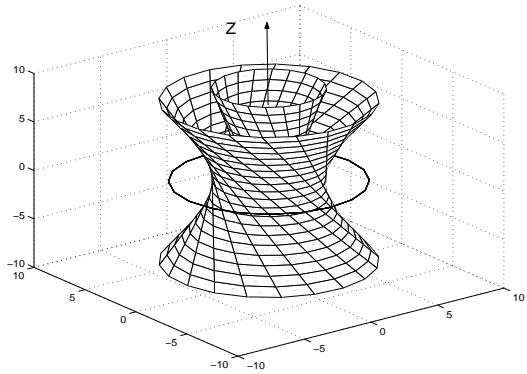,height=5cm,width=7cm}} \caption{Null directions of the Kerr
congruence $k^\m$ are focused on the Kerr singular ring, forming two-sheeted space of the advanced
and retarded fields.}
\end{figure}

\begin{figure}[ht]
\centerline{\epsfig{figure=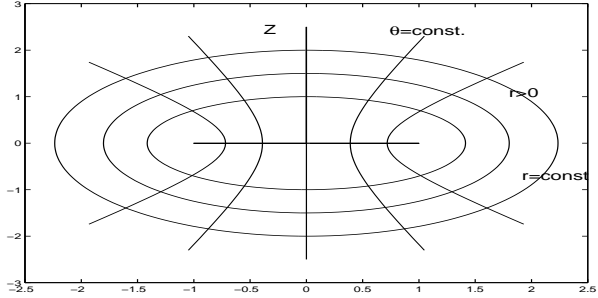,height=4cm,width=8cm}} \caption{Kerr's oblate spheroidal
coordinates  cover space-time twice, for $r>0$ and $r<0$.}
\end{figure}
Development of this model led in \cite{BurSol} to a soliton model with a domain wall phase transition, in which
Gravity controls external classical space-time, while the Quantum theory forms a supersymmetric
pseudo-vacuum state inside the bubble. The conflict between Quantum Theory and Gravity is resolved
by \textbf{the Principle of the separation of their zones of influence:}

\begin{description}
         \item[\textbf{PI:}] \textbf{\emph{space-time should be flat inside the core,}}
         \item[\textbf{PII:}]\textbf{ \emph{exterior should be exact KN solution}.}
         \item[\textbf{PIII:}]\textbf{ \emph{the boundary between regions PI and PII is determined by
         the L\'opez condition}} (\ref{Hre}).
       \end{description}

 It was mentioned in \cite{BurSol,Tomsk} a mysterious effectiveness of this principles which uniquely define form
 of this soliton and two its peculiarities:

\textbf{(A)}: the Higgs field is oscillating with the frequency $\omega= 2m ,$ and thus, it belongs a type of \emph{oscillons},

\textbf{(B)}: angular momentum is quantized,  $ J=n/2, \quad n=1,2,3,... \label{Jn/2} .$

\bigskip

In this paper we show that the KN buble-source forms a BPS-saturated soliton, and the both peculiarities
\textbf{(A), (B)} are uniquely determined by the Bogomolnyi equations, which determine also the shape of the soliton, and therefore,  its dynamics and stability.

Starting in sec.2 from description of our approach used in previous paper \cite{BurGrBag},
we derive in sec.3 the Bogomolnyi equations adapted to specific Kerr's coordinates, and integrate them, reducing the problem to two dimensions $(t,r) ,$ time and the Kerr radial coordinate.

In sec.4. We generalizing the stationary KN bag to the bag model flexible to deformations  and obtain that these deformations are also controlled with the Bogomolnyi bound. Considering stringy deformations of the bag caused by electromagnetic excitations of the KN solution, we show that traveling waves may create deformations which break smoothness of the solution and create a traveling singular pole connected with traveling circular wave.

Sec.5. is Conclusion.

\bigskip

\section{Gravitating bag model and supersymmetric scheme of phase transition} Bubble-source formed by
L\'opez boundary was generalized to soliton \cite{BurSol}), and then  to a gravitating bag model
\cite{BurGrBag,BurEmerg}. Conception of the bag model assumes incorporation of the fermionic sector
in which the Dirac equation acquires mass through Yukawa coupling with the Higgs field
\cite{MIT,SLAC}. As a consequence, the mass turns out to be a variable function of the space-time
distribution of the Higgs condensate. Boundary of the bag is modelled by a domain-wall
interpolating between  external KN solution  and flat internal  pseudo-vacuum state, and  phase
transition between these states is controlled by the Higgs mechanism of symmetry breaking, which is
used in many soliton models, as well as in the well-known Nielsen-Olesen model \cite{NO}, which is
in fact the Landau-Ginzburg (LG) field model for the vortex string in a superconducting media.

As it was shown in \cite{BurGrBag}, the typical quartic potential  $\Phi ,$ \be V(|\Phi|)=g(\bar\sigma
\sigma - \eta^2)^2 , \ \  \sigma =<|\Phi|> , \label{phi4} \ee used for the Higgs field in all these models,
is not suit for the source of the Kerr-Newman  solution, since the external Higgs field distorts the external
KN solution,  turning the electromagnetic field in the short-range one.

Contrary to the standard bag model forming  \emph{a cavity in the Higgs condensate}, \cite{MIT}, the
  condition $\textbf{PII}$ requires the Higgs condensate to be enclosed \emph{inside the bag.}
  This cannot be done with potential (\ref{phi4}), and there was used a more complex  scheme of
  the phase transition in \cite{BurGrBag}, which contained three chiral fields
$\Phi^{(i)}, \ i=1,2,3 .$ In fact it is a supersymmetric generalization of the LG model \cite{HLosShifm}.

  One of the fields, say $\Phi^{(1)} ,$ was identified as the Higgs field
$\Phi .$  Thus,  there were used new notations \be (\Phi, Z, \Sigma) \equiv (\Phi^1, \Phi^2, \Phi^3) . \ee

Due to condition $\textbf{PI} , $ bag is to be placed in the flat region, and domain wall phase transition
may be considered with the flat background  metric, $g_\mn = \eta_\mn .$ Therefore the domain wall boundary of the
bag and the bag as a whole are not dragged by rotation. Because of that, the chiral part of the
Hamiltonian  is simplified to \be H^{(ch)} = T_0^{ ~0  (ch)} = \frac 12 \sum_{i=1}^3 ~ [
\sum_{\m=0}^3  |\cD^{(i)}_\m \Phi^i|^2 +  |\d_i W|^2 ], \label{HamCh} \ee where the covariant
derivatives $\cD^{(i)}_\m \equiv \d_\m -ie A^i_\m $ are flat.  Following \cite{WesBag}, the potential
$V$ is determined by superpotential \be V(r)=  \sum _i |\d_i W|^2 . \label{VdW}\ee

It was shown in \cite{BurGrBag} that
the suggested by Morris \cite{Mor} superpotential
\be W(\Phi^i, \bar \Phi^i) = Z(\Sigma \bar \Sigma -\eta^2)
+ (Z+ \m) \Phi \bar \Phi,\label{W} \ee where $ \m$ and  $ \eta $ are the real constants, provides the necessary
concentration of the Higgs field inside the bag, and from the supersymmetry condition  $\d_i W =0 ,$ there were
determined two vacuum states:

\textbf{(I)} internal: $r<R-\delta $,
\be V (r) = 0 , \ |\Phi| = \eta = const., \ Z=-\m, \ \Sigma
=0, \ W_{in}=\m \eta^2 , \ee and

\textbf{(II)} external: $r>R +\delta $, \be V (r) = 0 , \ \Phi =0, \ Z=0, \ \Sigma=\eta, \ W_{ext}
=0, \ee and also

\textbf{(III)} Transition zone $R-\delta < r <R-\delta ,$ where the vacua \textbf{(I)} and \textbf{(II)} are separated by a positive spike of potential $V$.

The obtained here principal result is that the position of the domain wall boundary satisfying to requirements \textbf{PI-PIII}  is uniquely determined by Bogomolnyi bound, and therefore, these requirements determine stability of the bag, leading to a supersymmetric and BPS-saturated source of the KN solution.

As it was discussed in \cite{BurGrBag} (and earlier in \cite{BurSol}), inside the bag and in the transition zone (III) the space is flat,
 the fields $\Phi^2$ and $\Phi^3$  are constant, and only the complex  Higgs field $\Phi(x) = |\Phi (x)| e^{i\chi(x)} ,$
interacting with the penetrating inside vector potential of the KN solution $A_\m $ has a
nontrivial dynamics. As a result, the field model in this zone is reduced to  Abelian field model
in flat space-time, which has only one chiral field $\Phi$ and conincides with the used by
Nielsen-Olesen (NO) model for the vortex string in superconducting media \cite{NO}. The
corresponding Lagrangian leads to equations
\bea \d_\n \d^\n \Phi &=& \d_{\bar\Phi} V  , \label{PhiIn} \\
\d_\n \d^\n A_\m = I_\m &=&  e |\Phi|^2 (\chi,_\m + e A_\m), \label{Main} \eea
which are consistent with the vacuum states in zones (I) and (II).

The second equation (\ref{Main}), which is indeed the
eq. (2.4) of the NO model \cite{NO},  indicates that current must not penetrate inside the bag beyond
a thin surface layer. Putting inside the bag $ I_\m=0 ,$ we obtain $\d_\n \d^\n A_\m =0 $ and
\be\chi,_\m + e A_\m=0 , \label{compens}\ee which shows that gradient of the phase of the Higgs
field $\chi,_\m$ must compensate the penetrating vector potential $A_\m $ of the KN field. It
should be emphasized that, \emph{although the KN gravitational field vanishes near the border of
the bag, its strong effect on the electromagnetic field is maintained.}  Since the KN vector
potential
  \be A_{\m} = - Re  \frac {e} { (r+ia \cos \theta)} k_\m
\label{Amus}  \ee is aligned to directions of the Kerr congruence $k_\m ,$ it must be dragged by
the Kerr singular ring even in the flat limit, see Fig.3.

   \begin{figure}[h]
\centerline{\epsfig{figure=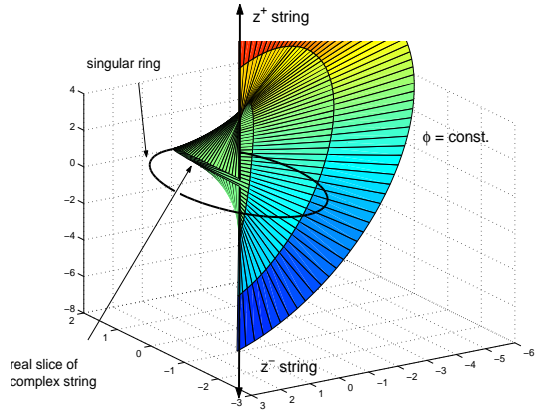,height=5.5cm,width=7cm}}
\caption{\label{label} The Kerr surface $\phi=const.$ The Kerr
congruence is dragged by rotation even in the limit zero mass. In the equatorial plane congruence is tangent
to the Kerr singular ring, and vector potential forms a closed Wilson loop wrapped along the border of the spheroidal bag.}
\end{figure}

 The boundary of the bag at $r=R=e^2/2m $ regularizes the vector potential (\ref{Amus}), and  it takes  the maximal value in the equatorial plane ($\cos \theta =0 $),
   \be A_{\m}^{(max)} = -  \frac {2m} { e} k_\m \label{Amax}.  \ee

There are only the longitudinal and the timelike components of the vector potential in
the stationary KN solution.
 Since $k_0 =1 ,$ the timelike component takes the maximal value $A_0 = - 2m/e ,$ which in accord with  (\ref{compens})
 should be compensated by phase of the Higgs field $\chi,_0 \ ,$ which leads to the important result \textbf{(A)}: oscillations of the Higgs field with the frequency $\omega= 2m .$

 In the same time, the longitudinal part of the vector potential $A_\m$ forms a closed loop along boundary of the bag
 in equatorial plane, and in the accord with (\ref{compens}) it should also be compensated by incursion of the
 phase of the Higgs field $\chi,_\phi \ .$ Using the Kerr relation $J=ma ,$ we obtained in \cite{BurSol}  the second remarkable consequence \textbf{(B)}: angular momentum is quantized,  $ J=n/2,
\quad n=1,2,3,... .$

\bigskip

\noindent We consider now these result as consequence of supersymmetry of the bag model. We use the recipe described in \cite{Cvet91,GibTown} for a similar problem for a planar domain wall with one chiral field and reduce the problem to the
solvable first order Bogomolnyi equations, obtaining in particular the consequences \textbf{(A), (B)}.

\section{Source of the KN solution as a BPS-saturated soliton}

The full Lagrangian corresponding to bosonic part of the N=1 supersymmetric model with three chiral fields $\Phi^{(i)} = \{\Phi, Z,\Sigma \},
 i=1,2,3,$  has the form \cite{WesBag} \be {\cal L}= -\frac
14 F_\mn F^\mn - \frac 12 \sum_i(\cD^{(i)}_\m \Phi^{(i)})(\cD^{(i)
\m} \Phi^{(i)})^* - V \label{L3} .\ee As we mentioned earlier, the part of the Lagrangian
which is related with field $\Phi^{(i)}=\Phi^{(1)}\equiv \Phi$ is the same as in the Nielsen-Olesen model.

The corresponding stress-energy tensor is decomposed into pure em part $T^{(em)}_\mn$
and contributions from the chiral fields $T^{(ch)}_\mn$, \be T^{(tot)}_\mn
= T^{(em)}_\mn +  \sum_i(\cD^{(i)}_\m \Phi^i)\overline
{(\cD^{(i)}_\n \Phi^i)} - \frac 12 g_\mn[\sum_i(\cD^{(i)}_\lambda \Phi^i)\overline {(\cD^{(i)\lambda} \Phi^i)}
+V] . \label{T3}  \ee.

 Flatness of the metric inside the bubble and in the vicinity of the domain
wall boundary leads to disappearance of dragging of the chiral fields, and
similar to previous treatment,  we can use the chiral part of the
Hamiltonian in the form (\ref{HamCh}).

The domain
wall boundary of the bag and the bag as a whole do not rotate. Nevertheless, the influence of
gravity is saved in the shape of the bag and also as a drag effect acting of the KN electromagnetic
field, which retains correlation with a twisted Kerr congruence even in the limit of the flat
space. We have to take it into account, and it is advisable to use the Kerr coordinate system
 \be x +iy =(r +ia) e^{i\phi} \sin \theta, \quad z= r\cos \theta , \quad t =\rho - r  \label{obl coord}, \ee
which is adapted with the shape of the bag, and where the KN vector potential (\ref{Amus}) takes the simple form (\cite{DKS}, eq.(7.7)) \be A_\m
dx^\m = - Re \ [(\frac e {r+ia \cos \theta})] (dr - dt - a \sin ^2 \theta d\phi ) \label{Am}. \ee

As we have seen, the components $A_\phi $ and $A_t $ have very specific behavior, and are compensated by the phase of the oscillating Higgs field \be \Phi (x) \equiv \Phi^1(x) = |\Phi^1 (r)| e^{i\chi(t,\phi)} , \label{Phichi}\ee  which is equivalent to  the equations
 \be \cD^{(1)}_t \Phi^1
=0, \quad \cD^{(1)}_\phi \Phi^1=0, \label{mainAH} \ee which are analogs of (\ref{Main}), and lead to consequences \textbf{(A)} and \textbf{(B)}
correspondingly.  As a result, these terms drop out from the expression  (\ref{HamCh}), and
all the remainder chiral fields depend only on the Kerr radial coordinate $r ,$
\be \Phi^2 = \Phi^2 (r), \quad \Phi^3 = \Phi^3 (r) . \ee
The sum $\sum_{\m=0}^3|\cD^{(i)}_\m \Phi^i|^2$ in (\ref{HamCh}) reduces  to the term

\be H^{(ch)} = T_0^{ ~0  (ch)} = \frac 12
\sum_{i=1}^3 ~ [ |\cD^{(i)}_r \Phi^i|^2 +  |\d_i W|^2 ] , \label{HamCh1} \ee

 where the coordinate $r$ parametrizes the  oblate  surface of the bag and, similar to parallel  surfaces of the planar domain walls, the surfaces $r$ and $r+dr$ may be considered as ``locally parallel'' to each other, see Figs.4,5.

\begin{figure}[ht]
\centerline{\epsfig{figure=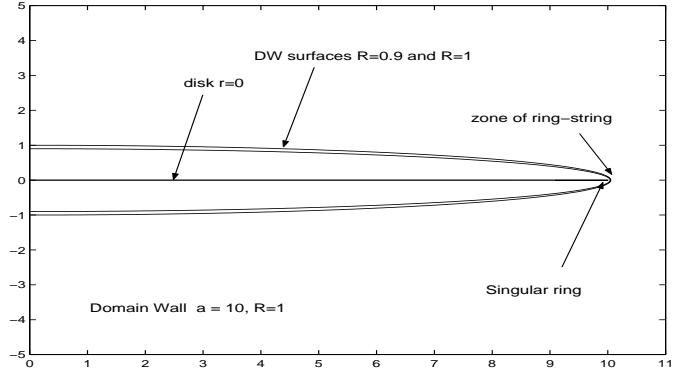,height=5cm,width=9cm}}
 \caption{Axial section of the spheroidal domain wall phase transition.}
\end{figure}

\begin{figure}[ht]
\centerline{\epsfig{figure=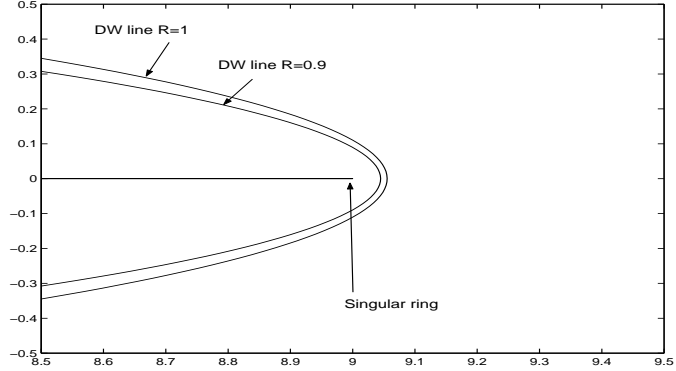,height=5cm,width=9cm}} \caption{Enlarged fragment of the disk border, the ring-string zone.}
\end{figure}

Following \cite{Cvet91,GibTown}, we use now a `trick', by introducing the angles $\alpha_i ,$ which allow us to  rewrite the expression  (\ref{HamCh1}) in the equivalent form
\be H^{(ch-r)} = \sum_{i=1}^3 \frac 12 | \cD^{(i)}_r \Phi^i - e^{i\alpha_i}\d \bar W /\d
\bar \Phi^i |^2 +  Re \ e^{-i\alpha_i}(\d \bar W /\d \bar \Phi^i) \cD^{(i)}_r \Phi^i ,
\label{HamCh2} \ee where the phases $\alpha_i $ should be independent on $r $ and be chosen to ensure
the vanishing of the square terms, i.e. \be \cD^{(i)}_r \Phi^i = e^{i\alpha_i}\d \bar W /\d \bar
\Phi^i . \label{Bog}\ee

The function $ W $ and $ Z $ are real, and without loss of generality, we can also set a real $\Phi^3 ,$ which allows us to take $\alpha_2 =\alpha_3 =0 .$  For the Higgs field, presented by function $\Phi \equiv \Phi^1 = |\Phi| e^{i\chi (t,\phi)} ,$ we have
$\frac {\Phi^1} {\bar\Phi^1} = e^{2i \chi (t,\phi)} ,$ and from  (\ref{Bog}) and (\ref{Phichi}) we obtain
\be \alpha_1 = 2\chi (t,\phi) .\ee  In this case $ \cD^{(1)}_r \Phi = e^{2i \chi (t,\phi)}  \cD^{(i)}_r \bar\Phi ,$ and
(\ref{HamCh2}) takes the form
\be H^{(ch-r)} = \sum_{i=1}^3 \frac 12 | \d_r \Phi^i - \d W /\d
 \Phi^i |^2 +  Re \ (\d  W /\d  \Phi^i) \d_r \Phi^i ,
\label{HamChd} \ee where the replacement of the covariant derivatives $\cD^{(1)}_r$ to partial
$\d_r$ is valid due to concrete form of the used superpotential (\ref{W}).

Minimum of the energy density $H^{(ch-r)}$ is achieved for
 \be \cD^{(i)}_r \Phi^i = \d  W /\d \Phi^i \quad \cD^{(i)}_r \bar\Phi^i = \d \bar W /\d \bar
\Phi^i , \label{Bog1}\ee which are the Bogomolnyi equations corresponding to saturated Bogomolnyi bound.
The expression (\ref{HamChd})  turns into full differential

\be H^{(ch-r)} =  Re \ (\d  W /\d  \Phi^i) \d_r \Phi^i =  \d W /\d r   . \label{HamChd1} \ee

We can now obtain the mass-energy of  the bag together with its  domain wall boundary

\be M_{bag} \equiv M_{ch}= \int dx^3 \sqrt{-g} \ T_0^{\ 0
(ch)} . \ee For the Kerr coordinate system  \be \sqrt{-g}= (r^2 +a^2 \cos^2 \theta) \sin \theta .\ee
 Axial symmetry allows us to integrate over $\phi ,$ leading to
\be M_{bag}= 2\pi \int dr d\theta  (r^2 +a^2 \cos^2 \theta) \sin \theta \ T_0^{\ 0 (ch)} .
\label{MbubKerr}\ee
Using (\ref{HamChd1})  we obtain
\be M_{bag}= 2\pi \int dr d\theta  (r^2 +a^2 \cos^2 \theta) \sin \theta  \d_r W    .
\label{MbagKerr}\ee
Taking into account that superpotential  $W (r)$ is constant inside and outside the source,
 \be W_{int} = \m\eta^2, \ W_{ext}=0, \ee
we have $\d_r W =0 $ inside and outside the bag and, by crossing the bag boundary, we get the incursion
$\Delta W = W(R+\delta) -  W(R -\delta) =  -\m \eta^2 $ .   After integration over $r\in [0,R] ,$ and then over $X=\cos\theta$
we obtain
\be M_{bag}= 2\pi \Delta W \int_{-1}^{1}  dX  (R^2 +a^2 X^2 ) = 4\pi(R^2  +\frac 13 a^2) \Delta W .\label{MbagKN}\ee

\section{Stringy deformations of the KN bag} As it was discussed in \cite{BurGrBag}, taking the bag
model conception, we should also accept the dynamical point of view that the bags are to be soft and deformed acquiring excitations, similar to excitations of the dual string  models \cite{SLAC,Giles,Vinci}. By deformations the bags may form stringy structures.
Generally considered the radial and rotational excitations, forming the open strings or flux-tubes. The old Dirac's model of an "extensible" spherical electron
\cite{DirBag} may also be considered as a prototype of the bag model with spherically symmetric deformations -- radial excitations.

The bag-like source of the KN solution without rotation, $a = 0 ,$  represents the Dirac model of a
spherical "extensible" electron, which has in rest the classical electron radius $R=r_e=e^2/2m .$
The KN rotating disk-like bag (see Fig.1. of the previous paper \cite{BurGrBag}), may be considered as the Dirac bag stretched by rotation
to a disk of the Compton radius, $a= \hbar/2mc ,$ which corresponds to the zone of vacuum
polarization of a ``dressed'' electron.

\begin{figure}[ht]
\centerline{\epsfig{figure=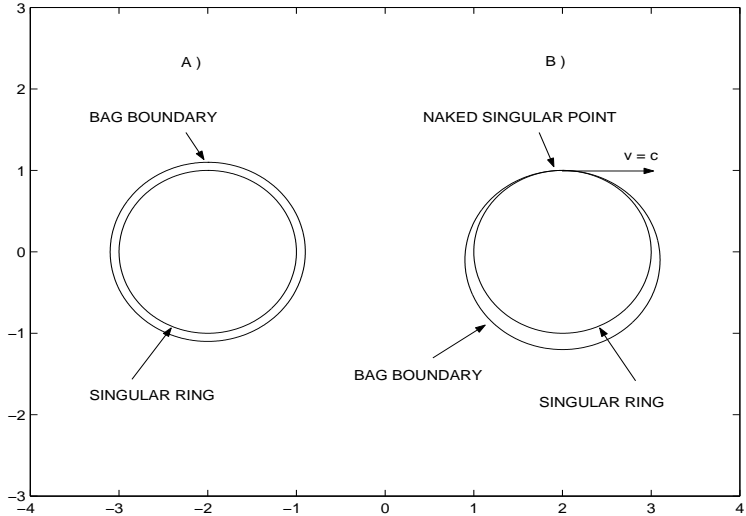,height=7cm,width=10cm}} \caption{Regularization of the KN
EM field. Section of the disk-like bag in equatorial plane. Distance from positions of the boundary
of the bag from position of the (former) singular ring acts as a cut-off parameter R. (A)Axially
symmetric KN solution gives a constant cut-off $R=r_e.$ (B)The boundary of the bag is deformed by a
traveling wave, creating a circulating singular point of tangency (zitterbewegung).} \label{fig4}
\end{figure}

It has been obtained long ago that the Kerr geometry is close related with strings \cite{Bur3Q}.
In particular, in our old work \cite{Bur0,IvBur}, the Kerr singular ring was associated with a closed
ring-string which may carry traveling waves like a waveguide.\fn{Another, complex string appears in the
 complex structure of the Kerr geometry \cite{Bur3Q,BurTMP}.}
In the soliton-bag model the Kerr singularity disappears, but this role is played by the
sharp border of the disk-like bag. Like the Kerr singular ring \cite{Bur0}, it can serve as carrier of
the traveling waves.
It was shown in \cite{BurSen} that field structure of this string is similar to the structure of the
fundamental string, obtained by Sen as a solitonic string-like solution to low energy string theory \cite{Sen}.
As it was shown still in \cite{Bur0,IvBur} and recently in \cite{BurHetStr},  the EM and spinor excitations of the KN solution are concentrated
 near the Kerr ring, forming the string-like traveling waves. For the stationary KN solution the EM field forms a
 \emph{frozen} wave \cite{Bur0}, located along the border of the disk-like source.  Locally, this  frozen string is the typical plane-fronted EM wave with null invariants
 \be (\mathbf{E}\cdot\mathbf{H}) =0, \ \mathbf{E}^2 - \mathbf{H}^2 =0,\ee and with the Poynting
 vector $\mathbf{S}=\frac 1{4\pi}[\mathbf{E}\times\mathbf{H}]$ directed tangently to  the Kerr singular
 ring, $(\mathbf{k} \cdot\mathbf{S}) >0 .$
 In the regularized KN solution, the Kerr singular ring is regularized, acquiring  the cut-off parameter $R$,
 which for the axially symmetric KN solution is the constant $R=r_e,$ of (\ref{Hre}), see Fig.6A.

Since the null vector of the Kerr congruence $k_\mu$ is tangent to the Kerr singular ring, and since
$R<<a$ the ring-string at the border is almost light-like, and its structure is very close to the known pp-wave strings
\cite{Dabhol,HorowSt,CvetTze}.
However, for an external observer, the light-like closed string should shrink to point due to Lorentz contraction, \cite{BurHetStr}.
The extended  KN string, positioned along the border of the bag, cannot be closed, \cite{BurOri}, since the end points of the string worldsheet
$x^\mu (\phi, t)$ and  $x^\mu (\phi+2\pi, t)$ must not coincide.\fn{Otherwise the worldsheet becomes a worldline. We are faced here with an odd peculiarity of the Kerr spinning particle, where the chiral fields form an extended bag, while the associated EM field
forms a lightlike string, size of which is reduced for any external observer.} There are two ways to make a consistent extended string structure:

1) to consider this string as an open one and to complete it to a consistent sum, comprising the left add right modes,

2) to form an orientifold string, i.e. the open string is built from a closed one by folding its worldsheet, \cite{BurOri}:
the interval $\phi \in [0, 2 \pi] $ is represented as a half-interval  $\phi^+ \in [0, \pi] , $ doubled by the reversed half-interval $\phi^- \in [\pi, 2\pi] ,$ with setting $x^\mu (\phi^-, t) = x^\mu (2\pi - \phi^- , t) .$

  Here we shall follow the first way, and consider the above ``frozen'' solution as right mode of an excitation. We will
  complete  it by the left counterpart, which we find among other admissible excitations.
All exact solutions for the EM field on the Kerr background were obtained in \cite{DKS}, and they are defined by analytic function
$ A= \psi(Y,\t)/P^2 $ where $ Y=e^{i\phi} \tan \frac \theta 2 $ is a complex
projective angular variable, $ \t = t -r -ia \cos \theta $ is a complex retarded-time parameter and
 $P=2^{-1/2}(1+Y\bar Y) $ for the Kerr geometry at rest.
 Vector potential is determined by function $\psi$ as follows \cite{DKS}
\be  A_\m dx^\m \\
= - Re \ [(\frac \psi {r+ia \cos \theta}) e^3 + \chi d \Y ], \quad  \chi =
2\int (1+Y\Y)^{-2} \psi dY  \label{alpha} \ee
\begin{figure}[ht]
\centerline{\epsfig{figure=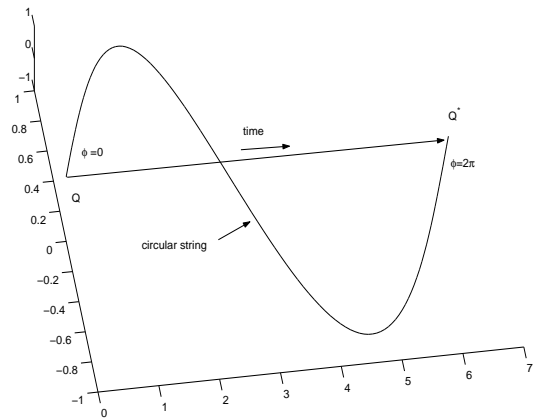,height=6cm,width=7cm}}
\caption{The circular left mode, formed by traveling wave along the KN string, is completed by the time-like right mode, formed by the frozen traveling wave of the stationary KH solution.}
\label{fig3}
\end{figure}
 The simplest function $\psi=-e $ yields the  stationary KN solution with function (\ref{HKN}). It corresponds to the discussed above
 frozen circular EM wave, see Fig.7. This circular traveling mode is locally plane wave  ``propagating''
 along  the Kerr singular ring.
 By regularization, the EM field gets the constant cut-off parameter $R=r_e$,
 see Fig.6A.

 Along with many other possible stringy waves, interesting effect shows  the lowest wave
 solutions\fn{Remarkable features of this combination were discussed still in \cite{Bur0}.}
 \be \psi = e(1 + \frac 1 Y e^{i\omega \t}) . \label{psi12}\ee
It is easy to find a back-reaction of this excitation.
 Boundary of the disk is very close to position of the Kerr singular ring, and
regularization of the \emph{stationary KN
 source} represents in fact a \emph{constant }cut-off parameter $R=r_e ,$ (\ref{Hre}) for the Kerr singularity.
 The EM traveling waves will deform the bag surface,
 and the boundary of the deformed bag can be determined from the condition $H=0 ,$ (\ref{HR0}).

 Like the stationary KN solution, function $\psi$ acts on the metric through the function $H$, which
 has in general case the form
 \be H =\frac {mr - |\psi|^2/2} {r^2+
a^2 \cos^2\theta} \ , \label{Hpsi} \ee and the condition $H=0$ determines the boundary of disk
$R=|\psi|^2/2m ,$ which acts as the cut-off parameter for EM field. The corresponding deformations
of the bag boundary are shown in Fig.6B. One sees that solution (\ref{psi12}) takes in equatorial
plane $\cos \theta=0$  the form $\psi = e(1 + e^{-i(\phi - \omega t)}) ,$ and the cut-off parameter
$R=|\psi|^2/2m =\frac {e^2}{m} (1 + \cos (\phi -\omega t)$ depends on $\phi - \omega t .$ Vanishing
$R$ at $\phi = \omega t$  creates singular pole which circulates along the ring-string together with
traveling wave of the excitation, reproducing the lightlike zitterbewegung of the Dirac electron.
This pole may be interpreted as a single end point of the ring-string, or as a point-like bare electron,
either as a light-like quark, if it will also be present in the associated fermionic sector.
\section{Conclusion}
The mysterious problem of the source of two-sheeted Kerr geometry leads to a gravitating soliton-bubble model,
which has to retain the external long-range  gravitational and EM field of the KN solution.
The requirement of consistency with gravity leads to a supersymmetric field model of phase transition,
in which the Higgs condensate forms a supersymmetric  core of a spinning particle-like solution.
The  considered in \cite{BurGrBag} resulting model has much in common with the famous MIT and SLAC bag models,
as well as with the basic conception of the Standard Model, where the initially massless leptons (left and
right) get a mass inside the bag from the Higgs mechanism of symmetry breaking.

In the present extension of the paper \cite{BurGrBag} we showed that the KN bag model forms a BPS-saturated solution
of the Bogomolnyi equations, and therefore, the stationary bag forms a stable configuration determined by the KN
parameters: charge, spin  and the rotation parameter $a =J/m ,$ while the mass is related with the parameters of the domain wall
bubble encoded in the superpotential $W.$

  Similar to the other bag models, the KN bag is pliant to deformations. Spinning bag takes the form of a thin disk,
  sharp border of which represents a ring-string, which can support the traveling waves. The domain wall boundary
of the disk is determined by BPS-bound which coincides with the L\'opez boundary determined by
principles \textbf{PI-PIII}. For the stationary KN solution it corresponds to the bag of the oblate
ellipsoidal form taking the Compton zone of a dressed electron. Boundary of the disk is completed by a ``frozen'' lightlike
ring-string of the Compton radius. Since tangent direction to this string is almost lightlike,  it shrinks by the
Lorentz contraction, \cite{ArcPer,BurHetStr}.\fn{However, it was supposed in \cite{ArcPer,BurHetStr} that the real Compton extension of this string would be
observable in some experiments with the low energy scattering.}

On the other hand, we showed that the ring-string traveling waves lead to deformations of the bag surface, and the lowest EM excitation of the KN solution breaks regularization of the KN solution creating a singular pole, which reproduces the known zitterbewegung, circulating with speed of light along the ring-string together with traveling wave. The bag model acquires an additional pointlike element which may be interpreted as an analog of the bare electron, while the model in whole turns into  a single bag-string-quark system, which should be associated with a dressed electron.

\section*{Acknowledgements}
Author would like to thank P. Kondratenko and Yu. Obukhov and all colleagues of the Theor. Phys.
Laboratory of NSI RAS for useful discussion. Author thanks also V. Dokuchaev, V. Rubakov and other
members of  Theoretical Division of INR RAS for invitation to seminar talk and useful discussion.
Author thanks J. Morris for reading a version of this paper, very useful conversation and correcting
some signs.  This research is supported by the
RFBR grant 13-01-00602.

\end{document}